\def \abc#1#2#3#4 {\reference#1, {\sl#2}, {\bf#3}, #4}
\def \blank {\lower 5pt\hbox to 0.75in{\hrulefill}}

\def \cm{~\rm{cm}}
\def \s{~\rm{s}}
\def \km{~\rm{km}}

\def \AU{~\rm{AU}}
\def \erg{~\rm{erg}}

\def \yr{~\rm{yr}}

\documentclass[12pt,preprint]{aastex}
\usepackage{natbib}

\begin{document}
%\normalsize
\small

\setcounter{page}{1}

%\shorttitle{COMMON ENVELOPE EVOLUTION}
%\shortauthors{SOKER}

%\slugcomment{Draft version of \today}

\title{ENERGY AND ANGULAR MOMENTUM DEPOSITION DURING
 COMMON ENVELOPE EVOLUTION}

\author{Noam Soker$^1$}

\affil{(1) Department of Physics, Technion$-$Israel Institute
of Technology, Haifa 32000, Israel;
and Department of Physics, Oranim; 
soker@physics.technion.ac.il}

\begin{abstract}
I consider three processes which enhance mass loss rate from a
common envelope of a giant star with a main sequence or
a white dwarf companion spiraling-in inside its envelope.
I consider deposition of orbital energy and orbital
angular momentum to the giant's envelope,
and in more detail the formation of jets by
an accreting companion and their propagation in the envelope.
I find that in many cases the deposition of orbital angular momentum
to the envelope may be more important to the mass loss process
than the deposition of orbital energy.
Jets blown by an accreting companion, in particular a white dwarf,
orbiting inside the outer regions of the giant's envelope may also
dominate over orbital energy deposition at early stage of the
common envelope evolution.
These imply that studies which ignore the deposition of angular
momentum to the envelope and the effects of the accreting companion
may reach wrong conclusions.
\end{abstract}

% ===================================================
\section{INTRODUCTION} 
% ===================================================

As a star in a binary system swells to become a giant it  
engulfs its companion if the orbital separation is smaller
than some critical value and if the companion is not
too massive; a common envelope (CE) phase commences.
(for a review see Iben \& Livio 1993, and Taam \& Sandquist 2000 ). 
Because of tidal interaction and friction,
the orbit shrinks. 
Several parameters can be defined to characterized the CE evolution
(e.g. Livio \& Soker 1988), but the most commonly used parameter is
the ratio of the binding energy of the ejected envelope
$\Delta E_{\rm bind}$ to the orbital energy that is released
during the CE phase $\Delta E_{\rm orb}$:
$\alpha_{\rm CE} \equiv \Delta E_{\rm bind}/\Delta E_{\rm orb}$.
Note that different definitions for the binding energy exist 
(e.g., O'Brien, Bond, \& Sion 2001).
Since the orbital energy that is released depends mostly on the final
orbital separation, the value of $\alpha_{\rm CE}$ can be in principle
calculated for systems whose final orbital separation is known,
assuming the giant structure at the onset of the CE is known
(O'Brien et al.\ 2001; Maxted et al.\ 2002).
The use of the $\alpha _{\rm CE}$ is common also in 
numerical simulations of the CE phase 
(e.g. Sandquist, Taam \& Burkert 2000 for a recent paper).
However, numerical simulations can't include the effect of enhanced
mass loss rate from giant stars that have a very high  mass loss rate.
The spun-up envelope of red giant branch (RGB) and asymptotic giant
branch (AGB) stars may have a much higher mass loss rate, with the
energy source being the giant's luminosity rather than the orbital
energy (Soker \& Harpaz 2003).

In some systems the usage of the above expression in a simple
manner yields $\alpha_{\rm CE} >1$.
For example, Maxted et al.\ (2002) assume that negligible mass has been 
lost prior to the onset of the CE phase in PG1115+166, and
find $\alpha_{\rm CE} >1$. 
This led some researchers to argue that the energy stored in the
envelope, and in particular the ionization energy, i.e., the energy
released when the envelope material recombines, is the extra energy needed
to expel the CE (e.g., Han, Podsiadlowski, \& Eggleton 1994;
Dewi \& Tauris 2000; Maxted et al.\ 2002). 
This proposed mechanism was criticized in previous papers
(Harpaz 1998; Soker 2002; Soker \& Harpaz 2003). 
In Soker (2002) I criticized the paper by Maxted et al.\ (2002) for not 
considering the mass lost from the envelope prior to the onset of the CE,
when the system is synchronized, i.e., the giant's rotation period
equals the binary orbital period, and the binary orbital shrinkage
proceeds very slowly. 
Eggleton (2002), for example, notes that a close companion may 
substantially enhance mass loss rate prior to the onset of a 
Roch lobe overflow (RLOF), with the possibility of preventing 
a CE phase altogether.

Soker \& Harpaz (2003) criticize Han et al.\ (2002)
for claiming that the ionization energy in the envelope is a significant
factor in the CE evolution. 
Soker \& Harpaz (2003) consider the mass lost by RGB stars
as they expand by a relatively large factor from the moment of
synchronization to the RLOF.
Soker \& Harpaz then argue that Han et al.\ (2002) include a mass loss
rate prior to the onset of the CE that is too low, and do not 
include the energy radiated by the accreting white dwarf companion, 
as well as that emitted by the core of the giant star.
In a later paper Han et al.\ (2003) briefly refer to Soker \& Harpaz 
criticism, keeping the dispute alive. 
Since the applicability of the $\alpha_{\rm CE}$ parameter is a fundamental
question in the CE process, and the CE evolution is the channel for the 
formation of many close binary systems, I elaborate on some questions 
regarding energy and angular momentum budget in the CE phase. 
In the first several sections I study the way by which an accreting
companion can deposits energy via jet formation.
I then (section 7) put all into a coherence picture.
A short summary is in section 8. 

% ===================================================
\section{MASS ACCRETION RATE}
% ===================================================

The Bondi Hoyle mass accretion rate inside the envelope is
(Armitage \& Livio 2000)
\begin{equation}
\dot M_{\rm acc} =  \pi \left( \frac{2 G M_2}{v_r^2 + C_s^2} \right)^2
\rho_e ( v_r^2 + C_s^2 )^{1/2},
\end{equation}
where $M_2$ is the mass of the accreting companion, $v_r$ is the
relative velocity of the accreting companion relative to the
unperturbed envelope, $\rho_e$ is the unperturbed envelope
density at the location of the accreting star, and $C_s$ is
the sound speed inside the unperturbed envelope.
The companion orbits the giant's core at the Keplerian velocity
$v_K$.
However, the relative velocity $v_r$ will be somewhat smaller
because the envelope is likely to be spun-up by the spiraling-in
companion.
Considering also that the motion inside the envelope is mildly
supersonic (Armitage \& Livio 2000), I use the approximation
$(v_r^2+C_s^2)^{1/2} \simeq v_K$.
The difference in the value of the accretion rate as a
result of this approximation will be absorbed in the parameter
$\zeta$.
For an AGB envelop density a good approximation for the
present purpose is (Soker 1992)
\begin{equation}
\rho_e= \frac {M_{\rm env}}{4 \pi R_\ast}
\frac {1}{r^2},
\end{equation}
where $M_{\rm env}$ is the envelope mass and $R_{\ast}$ the
stellar radius.
For the formation of a CE the companion shouldn't bring the envelope
to corotation, hence $M_2 \lesssim 0.3 M_1$. As it spirals-in,
the mass inward to the secondary orbit, $M_1(a)$ decreases,
and the mass ratio can become as large as $M_2/M_1(a) > 1$.
Here $M_1(a)=M_c+M_{\rm env}a/R_\ast$ is the giant mass inward to
the companion location, $a$ is the distance of the companion from the
core, and $M_c$ is the giant's core mass.
However, to draw my main conclusions I am interested in the
accretion process in the outer regions of the envelope,
where a crude but adequate approximation can be 
$M_1(a) \gg M_2$.
The Keplerian velocity is then simply $v_K=[GM_1(a)/a]^{1/2}$.
By using this and equation (2) in equation (1), I derive
the accretion rate in the form
\begin{equation}
\dot M_{\rm acc} \simeq 2 \pi \zeta 
\frac{M_{\rm env}}{\tau_K}
\left[\frac{M_2}{M_1(a)} \right]^2
\frac{a}{R_\ast},
\end{equation}
where $\tau_k=2 \pi a/v_K(a)$ is the Keplerian orbital period.
 For a WD with $M_2 \sim 0.6 M_\odot$, inside an AGB stellar envelope
with $M_{\rm env} \simeq 1 M_\odot$,
$M_c \simeq 0.6 M_\odot$, $R_\ast \simeq 1-2 \AU$, 
I find $\dot M_{\rm acc} \sim \zeta M_\odot \yr^{-1}$ in most of 
the envelope.
This is a too high an accretion rate as it exceeds the Eddington limit
of
\begin{equation}
\dot M_{\rm Edd} = 4 \pi m_p c R_2 \sigma_T^{-1}
= 10^{-3} 
\frac{R_2}{R_\odot} M_\odot \yr^{-1},
\end{equation}
where $R_2$ is the radius of the accreting star, $m_p$ the proton mass,
$c$ the speed of light, and $\sigma_T$ the Thomson cross section.
I find that $\zeta \sim 10^{-3}$ and $10^{-5}$, for
main sequence stars and WDs, respectively.
As the accreting star swells, the accretion rate increases,
possibly leading for further expansion of the secondary's envelope.

% ===========================================================
\section{ANGULAR MOMENTUM ACCRETION RATE}
% ===========================================================

I turn now to consider the specific angular momentum of the
accreted mass.
For a density gradient perpendicular to the relative velocity
of the ambient medium and the accreting mass (the $y$ direction)
of $\rho=\rho_0(1+y/H)$, the specific angular momentum 
of the accreted mater is
\begin{equation}
j_{\rm acc} = \frac{\eta}{4 H} 
\frac{ (2 G M_2)^2}{v_r^3},
\end{equation}
where $\eta\sim 0.25 $ is the ratio of the accreted angular momentum to
that entering the Bondi-Hoyle accretion cylinder, and it depends on
the Mach number and the equation of state (Livio et al.\ 1986).
For the envelope density profile given in equation (2) $H=r/2$,
but at early AGB phases, before much of the envelope has been lost and
the star did not reach its full size on the AGB,
it is steeper with $ H \simeq r/4$; I take $H=r/2$, as it is
not a bad approximation, and it also fits a wind outside the envelope.
 Taking again the Keplerian velocity for the relative velocity,
I find
\begin{equation}
j_{\rm acc} \simeq 2 \eta \left[ \frac {M_2}{M_1(a)} \right]^2
[G M_1(a) a]^{1/2}.
\end{equation}
If $\zeta \ll1$ and the mass is accreted with an impact parameter
smaller than the Bondi-Hoyle radius, then $\eta \ll 0.25$.
On the other hand, if a polar outflow is formed such that it
prevents some accretion from the polar directions, i.e., most of
the accreted mass comes from and near the equatorial plane, then
the specific accreted angular momentum will be higher. 
To form an accretion disk, $j_{\rm acc}$ should be larger 
than the angular momentum of a Keplerian motion on the 
companion's equator $j_2 = (GM_2R_2)^{1/2}$, 
where $R_2$ is the radius of the accreting companion.
 The condition $j_{\rm acc} > j_2$ gives
\begin{equation}
2 \eta \left[ \frac {M_2}{M_1(a)} \right]^{3/2}
\left( \frac {a}{R_2} \right)^{1/2} \gtrsim 1 .
\end{equation}
For the formation of a CE,
$M_2/M_1 \lesssim 0.3$, otherwise the companion brings the envelope
to corotation. Inside the envelope, where $M_1(a)<M_1$, this ratio
becomes larger.
Condition (7) then reads $a \gtrsim 100 R_2$.

%=====================================
\section{A MAIN SEQUENCE COMPANION}
%=====================================
From equation (7) it can be seen that an accretion disk will not form
around a main sequence companion in a CE phase well inside
the envelope, while in the outer regions of the envelope an accretion
disk can be marginally formed.
When the companion enters the envelope, the density profile is
very steep, and for a short time the accretion rate of
specific angular momentum is very high; at this stage a temporary
accretion disk might be formed, if was not present before
i.e., due to Roche lobe over flow (RLOF) or accretion from a wind.
In any case, the mass accretion rate increases substantially as
the companion enters the envelope;
a short burst of two opposite jets may result from this phase.

Outside the envelop the wind's density also falls as $r^{-2}$
(for a wind with constant speed and mass loss rate), and
we can crudely use equation (7).
For a main sequence companion outside the envelope,
$M_2$ is larger and/or the separation must be such that no
tidal interaction occurs, i.e., $R \gtrsim 5 R_\ast$,
where $R_\ast$, as before, is the radius of the giant.
These requirements make the formation of an accretion disk
in detached systems much more likely
(if the orbital separation is not too large).

The simple treatment above leads to a strong conclusion. 
While a main sequence star outside an AGB star (or other giants)
may accrete and form an accretion disk (see Soker 2001 for
the conditions for that to occur), a main sequence star 
inside the giant envelope will form an accretion disk for a short
time. It is not clear if jets can be blown during this short time.
If jets, or collimated fast wind (CFW), are blown outside the envelope
a bipolar PN is formed.
The results may explain the observations that most PNe with close
binary nuclei are not bipolar PNs.
The exception in NGC 2346, which has the largest known
orbital period (Bond \& Livio 1990; Bond 2000).
In that system the onset of the CE probably occurred at
a late stage, and the companion blew the CFW while still
outside the envelope (Soker 2002).

We saw above that an accretion disk is unlikely to be formed around
a main sequence companion spiraling-in inside the envelope.
However, the accreted angular momentum spins-up the companion,
enhancing the magnetic activity of an accreting main sequence
companion with a convective envelope.
Enhanced magnetic activity of spun-up main sequence companions
which accrete from the winds of AGB stars was considered before,
for main sequence companions of WDs (Jeffries \& Stevens 1996), and
for companions of central stars of PNs (Soker \& Kastner 20002).
Here I consider the case of accretion inside the envelope rather
than from a wind.

During the short time of the CE phase and the
following PN phase, most of the accreted angular momentum will
be distributed in the outer convective layer of the accreting
main sequence companion.
Let the mass in this layer be $M_{\rm con}$, and its moment
of inertia $\delta M_{\rm con} R_2^2$.
Equating the angular momentum in that layer to the accreted one
gives an expression for the rotation period of the
spinning main sequence companion (or at least its
convective layer) $P_{\rm rot}$.
The accreted angular momentum is $j_{\rm acc} M_{\rm acc}$,
where $j_{\rm acc}$ is the specific accreted angular momentum
as given by equation (6), and $M_{\rm acc}$ is the accreted mass,
of order $0.01-0.05 M_\odot$ (Hjellming \& Taam 1991).
Neglecting the envelope angular momentum prior to accretion, gives 
\begin{equation}
\delta M_{\rm con} R_2^2 \frac {2 \pi}{P_{\rm rot}} \simeq
2 \eta M_{\rm acc} \left[ \frac {M_2}{M_1(a)} \right]^2 
[G M_1(a) a]^{1/2}.
\end{equation}
To an order of magnitude, I take for the average value
during accretion $a \sim 50 R_\odot$, and $M_1(a)=1 M_\odot$,
and obtained the following scaled expression for the rotation
period
\begin{equation}
P_{\rm rot} \simeq 5
\left( \frac {\delta M_{\rm con}}{5 M_{\rm acc}} \right)
\left( \frac {\eta}{0.2} \right)^{-1}
\left( \frac {R_2}{R_\odot} \right)^2
\left( \frac {M_2}{M_\odot} \right)^{-2}
{\rm hrs}.
\end{equation}
Since some angular momentum will be transferred to the inner
radiative envelope, the rotation period will be somewhat longer.

 A low mass main sequence companion is likely to end close
to the core, hence being strongly influenced by tidal forces,
as are main sequence companions in cataclysmic variables.
Main sequence companions to WDs in cataclysmic variables are
known to be magnetically active (Saar \& Brandenburg 1999,
and references therein);
Saar \& Brandenburg term them magnetically superactive stars,
and review their properties in relation to other active stars.
The typical rotation period is $\sim 2~$hours$-2~$days, and the
magnetic activity cycle period of these stars is $5-50 \yr$
(this is the activity cycle, i.e., for the Sun it is 10 years,
rather than the full 20 years cycle).
In a previous paper (Soker \& Livio 1994) I proposed that
a main sequence star emerging from a CE with an
AGB star may, under certain conditions, transfers mass to the
core (of the previous AGB star) at a rate of
$\sim 10^{-6} - 5 \times 10^{-5} M_\odot \yr^{-1}$.
If a disk is formed around the core, I proposed, then two
jets may be blown, with a mass loss rate into the jets of
$\sim 10^{-8} - 10^{-6} M_\odot \yr^{-1}$.
If the magnetic activity cycle regulate the mass loss rate from
the main sequence star to the core, then the jets will have
a periodic (or semiperiodic) density variation.

%=====================================
\section{A WHITE DWARF COMPANION}
%=====================================

The situation with a WD companion is  more complicated.
 Hachisu, Kato \& Nomoto (1999) examine a spherically symmetric
accreting WD.
 They find, for their prescribed mass loss rate, 
that a  WD accreting at a rate of (the Eddington luminosity
of a WD)
$\dot M_{\rm acc} \gtrsim 10^{-5} M_\odot \yr^{-1}$, 
swells substantially, up to several solar radii.
As its radius increases, the Eddington mass accretion rate increases,
and the WD can swell to tens of solar radii, itself becoming a giant.
However, the angular momentum of the accreted mass must be considered.
Unlike an accreting main sequence star, the envelope formed around
the WD is made completely of the accreted gas, hence has a large
specific angular momentum.
Let the radius of the swelled WD be $R_s$, the envelope mass
$M_{\rm acc}$, and its moment of inertia
be $I_s=k_s M_{\rm acc} R_s^2$, where $k_s \sim 0.2$.
Using equation (6) for the specific angular momentum of the accreted
matter, assuming that no angular momentum is lost, and
assuming that the WD envelope rotates as a solid body,
I derive the ratio of the WD envelope angular velocity to the
break-up velocity on its equator, i.e., the Keplerian angular
velocity on the equator $\omega_{\rm Kep}$, 
\begin{equation}
\frac{\omega_s}{\omega_{\rm Kep}}
= 0.8 \frac{\eta}{k_s}
\left( \frac {R_s}{10 R_\odot} \right)^{-1/2}
\left( \frac {a}{100R_\odot} \right)^{1/2}
\left[ \frac {M_2}{0.25M_1(a)} \right]^{3/2} .
\end{equation}
Some angular momentum will be lost, however, via mass loss,
as is the case in accretion disks. 
 This show that, under these assumption, the WD will form a fast
rotating envelope, which is highly deformed to an oblate shape,
possibly with `dips' along the poles, i.e., something similar
to a thick accretion disk.
Such an envelope may form collimated fast wind (CFW)
along the polar directions.

 A full two-dimensional numerical calculation is required to find the
fate of a WD accreting mass with high specific angular momentum.
 Here I only point to the possibility that the accreting
WD may blow jets.
The accreting WD has a strong energy source, the nuclear-burning
on its surface, which is
$L_{WD} \simeq 2 \times 10^4 L_\odot$ for a WD of mass $0.6-0.8 M_\odot$.
Typical for accretion disks is that $\sim 10 \%$ of the accreted mass
is blown in jets.
If $10 \%$ of the nuclear-burning energy goes to blow jets,
I can estimate the mass loss rate to the two jets, $\dot M_j$, from
$0.1 L_{\rm WD} = \dot M_f v_e^2/2$, where $v_e$ is the escape velocity
from the poles of the swollen WD.
 This gives,
\begin{equation}
\dot M_j \simeq 10^{-4} 
\left( \frac {R_s}{1 R_\odot} \right) M_\odot \yr^{-1},
\end{equation}
which holds if the accretion rate is
$\dot M_{\rm acc} \gtrsim 10^{-3} M_\odot \yr^{-1}$.

% ===================================================
\section{JET PROPAGATION} 
% ===================================================

I now consider the possibility that two not-well
collimated jets (i.e., a  CFW), one at each side of the equatorial
plane, are blown by an accreting WD (or a neutron star) companion.
Jets blown by the core of an AGB star, via the destruction
of a brown dwarf for example, can easily clean a path
inside the envelope and emerge on the poles (Soker 1996).
When the orbiting companion blows the jets, the jets need
to penetrate different regions along the orbit, hence
they are less likely to emerge on the surface.
The relevant properties of the highly supersonic jet
are its speed $v_j$, its half opening angle (from its
symmetry axis to its edge) $\theta \ll 1$,
and the mass loss rate into each jet
$\dot M_j = \beta \dot M_{\rm acc}$, where
$\beta$ is the fraction of the accreted mass blown into each jet.
The density inside the jet, which propagates perpendicular to
the orbital plane along the $z$ axis, is
\begin{equation}
\rho_j =  \frac {\beta \dot M_{\rm acc}} {\pi z^2 \theta^2 v_j}. 
\end{equation}
The envelope density, by equation (2), is
\begin{equation}
\rho_e= \frac {M_{\rm env}}{4 \pi R_\ast} \frac{1}{a^2+z^2},
\end{equation}
where $a$ is the orbital separation. 
The head of the jet proceeds at a speed $v_h$ given by the balance
of pressures on its two sides.
Assuming supersonic motion, this equality reads
$\rho_e v_h^2 = \rho_j (v_j-v_h)^2$.
Eliminating $v_j/v_h$, using equation (13) for $\rho_e$,
and equation (12) for $\rho_j$, with $\dot M_{\rm acc}$ from
equation (3), I derive the following expression
\begin{equation}
\frac {v_j}{v_h}-1 \simeq
\frac {z}{(a^2+z^2)^{1/2}} \frac{\theta}{2 (\zeta \beta)^{1/2}}
\left( \frac{v_j}{v_K} \right)^{1/2}
\frac {M_1(a)}{M_2}.
\end{equation}
Close to the jet's source, i.e., $z \ll a$, the jet's head proceed
at a speed close to $v_j$.
Further away it slows down, and because the jet is not well collimated,
as was deduced in previous sections, i.e., $\theta \gtrsim 0.2$,
we have $v_h \ll v_j$.
Neglecting therefore the term $`-1'$ in the last equation,
I find for $z \gtrsim a$
\begin{equation}
\frac {v_h}{v_K} \lesssim
\frac {(a^2+z^2)^{1/2}}{z} \frac{2 (\zeta \beta)^{1/2}}{\theta}
\left( \frac{v_j}{v_K} \right)^{1/2}
\frac {M_2}{ M_1(a)}.
\end{equation}
 As an example I take the following values:
 A jet from a WD with $v_j=3000 \km \s^{-1}$, Keplerian velocity on
the giant stellar surface of $v_K(R_\ast)=30 \km \s^{-1}$, $\theta=0.2$,
$\zeta=10^{-5}$, and a high efficiency of jet formation
at the Eddington limit $\beta=0.5$ (the mass which is blown in the
jets is equal to the accreted mass), and $M_2 \simeq 0.2 M_1(R_\ast)$.
At $z \simeq a$, and at $a \simeq 0.5 R_\ast$, equation (15) gives
$v_h \lesssim 0.1 (\theta/0.2)^{-1} v_K$. 
I find for these parameters $v_h \lesssim 10 \km \s^{-1}$.
For neutron star, with $v_j \simeq 10^5 \km \s^{-1}$,
$\zeta \sim 10^{-8}$and $\theta \simeq 0.1$, similar values are obtained.
By equation (11) a swollen WD can have a much higher mass loss rate,
but the jet speed will be lower, and the jets will be much less
collimated.
Over all, the jet's head will proceed at a subsonic speed inside
the envelope with $v_h \lesssim 30 \km \s^{-1}$.

I now examine the distance the jets propagate inside
the envelope before slowing down.
The total width of the jet at a distance $z$ from the equatorial
plane is $2 \theta z$.
The orbiting jet-blowing companion crosses this distance along
its orbital motion in a time of $t_c=2 \theta z /v_K$.
During that time the jets proceeds a distance of
\begin{equation}
\Delta z \simeq t_c v_h \simeq 
4 (\zeta \beta)^{1/2} (a^2+z^2)^{1/2}
\left( \frac{v_j}{v_K} \right)^{1/2}
\frac {M_2}{M_1(a)}.
\end{equation}
Note that the distance does not depend on the opening angle of the
jet.
For the parameters used in the previous example,
$\Delta z < 0.1  z$.
This means that the jet's head at a distance $z\sim a$ from
the equatorial plane, will move only a distance $\sim 0.1 z$ before
the supply of fresh jet's material ends (because the companion
moves along its orbit).
 I found above that the jet's head moves at a subsonic speed of
$v_h \lesssim 30 \km \s^{-1}$.
Hence, after the supply of fresh jet's material ceases, 
the jet will not propagate much. Even if exit the envelope,
its speed will be below the escape velocity from the envelope.

 These jets have a different large effect on the envelope.
The jets shocked to a very high temperature, and form a hot and
low-density bubble, which is then buoyant outward, and ,mechanically
disturbed the envelope.
It is true that the luminosity of the accreting star deposited
more energy, but it is thermal energy which the stellar envelope can
transport outward via the convective envelope
(small fraction of the radiated energy goes to accelerate the
wind of RGB and AGB stars).
The large bubble (see below) may have a larger effect on the envelope.
Let the energy deposited into the shocked jets be
$\sim \chi L_{\rm Edd}$.
For a mass loss rate into the two jest of
$\dot M_j = 0.05 \dot M_{\rm Edd}$ (as noted earlier, due to the
nuclear burning, the mass loss rate can be somewhat higher),
and jet speed equal to the Keplerian speed at the accreting
companion surface, we find $\chi =0.025$. 
 This energy forms a bubble, whose change of volume rate
is given by energy considerations
$\dot V \simeq f_b \chi L_{\rm Edd}/P(r)$, where $P(r)$ is the pressure
in the envelope, and $f_b =0.4$ for an adiabatic index of $\gamma=5/3$.
 For a $\sim 1 M_\odot$ envelope a good approximation is
 (Soker \& Harpaz 1999)
\begin{equation}
P(r) \simeq 10^5
\left( \frac {r}{100 R_\odot} \right)^{-3} \erg \cm^{-3}.
\end{equation}
During one orbit, of duration 
$\tau_K \simeq 0.3 ({r}/{100 R_\odot})^{3/2} \yr$, the ratio of the
volume filled by the shocked two-jets' material $V_K$,
to the volume inner to the orbit is, for
$L_{\rm Edd}=10^{38} \erg \s^{-1}$,
\begin{equation}
\frac {V_K}{4 \pi r^3 /3} \simeq 0.05 
\left( \frac {r}{100 R_\odot} \right)^{3/2} 
\left( \frac {\chi}{0.025} \right) .
\end{equation}
However, the orbit-decay time is longer than $\tau_K$
(Hjellming \& Taam 1991).
Substituting $1 \yr$ instead of $\tau_K$, gives
\begin{equation}
\frac {V (1 \yr)}{4 \pi r^3 /3} \simeq 0.2 
\left( \frac {\chi}{0.025} \right) .
\end{equation}
I find that the hot bubble formed by the two jets near orbit
of an accreting WD will cause a large disturbance in the envelope.
This may facilitate the ejection of the envelope in a CE evolution.

% ===================================================
\section{THE OVERALL PICTURE} 
% ===================================================

% ===================================================
\subsection{Momentarily Effects} 
% ===================================================

To put the mechanical energy of jets into the overall picture,
I consider 3 mechanisms as orbits shrinks.
The first is energy deposition.
From the energy (gravitational plus kinetic) of the orbiting binary
system, $E_{\rm orb}$, I find (I define positively deposited energy)
\begin{equation}
\frac {d E_{\rm orb}}{d a}
= \frac {G M_1 M_2} {2 a^2} .
\end{equation}
A fraction $1-\alpha_e$ of this energy be radiated away, as the nuclear
energy produced by the giant is, and will not be used in expelling
the envelope.
For an envelope density profile of $\rho \propto r^{-k}$,
with $k \simeq 2$, the envelope binding (negative of gravitational)
energy is
\begin{equation}
\Delta E_{\rm bind} = B_e \frac{G M_{\rm env} M_1} {R_\ast},
\end{equation}
where $B_{\rm env} \sim 5-10$.
The relative energy deposition as the companion spirals-in is
defined as
\begin{equation}
D_E (M_2,a) \equiv \frac {\alpha_e}{\Delta E_{\rm bind}}
\frac {d E_{\rm orb}}{d \ln a}
= \frac {\alpha_e}{2 B_{\rm env}} \frac {M_2}{M_{\rm env}}
\frac {R_\ast}{a} = 0.05
\frac {\alpha_e}{0.5} 
\left( \frac {B_{\rm env}}{5} \right)^{-1}
\frac {M_2}{M_{\rm env}}
\frac {R_\ast}{a}. 
\end{equation}
This quantity represents the relative importance of the
orbital energy deposited into the envelope as the orbit shrinks
by a short radial distance $da \ll a$.

The rate of orbital angular momentum deposited to the envelope as the
orbit shrinks is given by
\begin{equation}
\frac {d J_o}{d a}
= \frac {1}{2} \left[ \frac {G(M_1+M_2)}{a} \right]^{1/2}
\frac {M_1 M_2}{M_1+M_2}.
\end{equation}
The moment of inertia of the envelope, for the density profile 
assumed above, is $I_{\rm env} = k_e M_{\rm env} R_\ast^2$,
with $k_e \simeq 0.2$.
The maximum angular momentum of the envelope, assuming
a uniform rotation is then
\begin{equation}
J_{\rm env} ({\rm max}) 
= k_e M_{\rm env} (G M_1 R_\ast)^{1/2}.
\end{equation}
The relative importance of angular momentum deposition is defined
as
\begin{equation}
D_J \equiv \frac {1}{J_{\rm env} ({\rm max})} \frac {d J_o}{d \ln a}=
\frac {1}{2k_e}
\frac{M_1^{1/2}}{(M_1 + M_2)^{1/2}}
\frac{M_2}{M_{\rm env}}
\left(\frac {a}{R_\ast} \right)^{1/2}
\simeq 2 \frac{M_2}{M_{\rm env}}
\left(\frac {a}{R_\ast} \right)^{1/2} ,
\end{equation}
where I used $M_1 \gg M_2$ in the last equality.
Note that $D_J$ is not identical to the
$\gamma_{\rm CE}$ parameter defined as the envelope spinning-up
time scale to the orbital decay time scale (Livio \& Soker 1988),
although they are similar in representing the significance of
envelope spin-up.
Here $D_E$ and $D_J$ represent the effects of energy and angular
momentum deposition, respectively, as the orbit shrinks by
a small amount $d a$.

The mechanical energy deposited by the jets (including both the
$PdV$ work and the internal energy of the gas inside the bubble)
depends on several poorly known parameters, i.e., $\chi$ and
the time-scale for orbital decay $\tau_d = a/\dot a$.
Using the Eddington accretion rate from equation (4)
with its Eddington luminosity and the jets' mechanical energy
used in equations (18) and (19), and using the
envelope binding energy used above, the relative importance
of the mechanical energy of the jets is defined as
\begin{equation}
D_{\rm acc} \equiv \frac {\tau_d \chi L_{\rm Edd}}{\Delta E_{\rm bind}}
= 0.25 \left( \frac {B_{\rm env}}{5} \right)^{-1}
\left( \frac{M_1}{M_\odot} \right)^{-1}
\frac {R_\ast}{500 R_\odot}
\frac {\tau_d}{100 \yr}
\frac{\chi}{0.025} 
\frac {M_2}{M_{\rm env}}.
\end{equation}
The long decay time-scale used here is appropriate when the companion
is in the outer regions of the giant's envelope. 

In Figure 1 I plot the three functions, $D_E$, $D_J$, and $D_{\rm acc}$,
\begin{figure}
\plotone{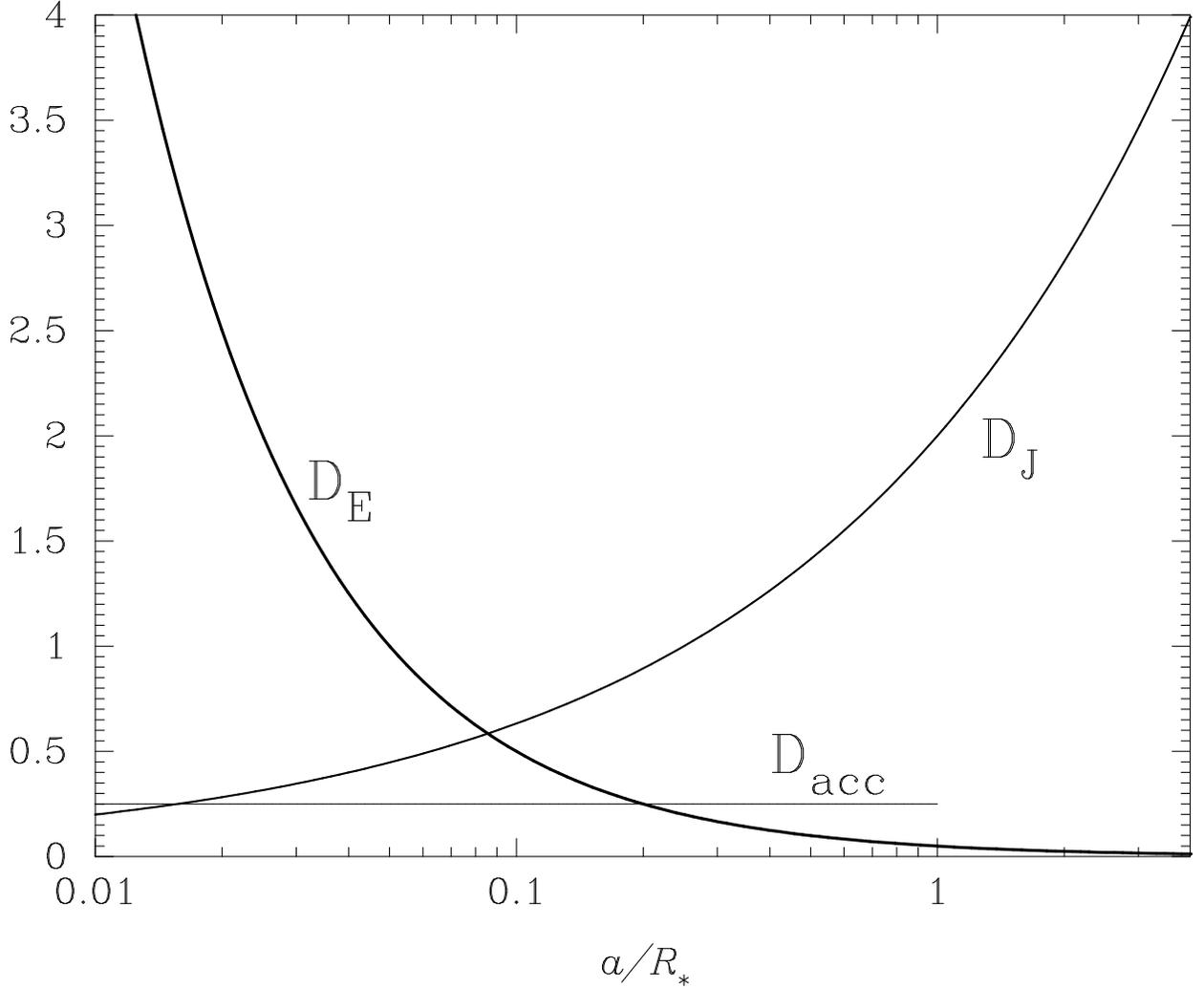}
\caption{The relative importance of the deposition of three quantities
to the common envelope as the orbit shrinks by a small amount
$da \ll a$, as a function of the orbital separation $a$, in units
of the giant's radius $R_\ast$. 
Drawn are the deposited orbital energy relative to the envelope
binding energy $D_{\rm E}$ (eq. 22), the deposited orbital
angular momentum relative to the maximum possible for the giant's
envelope $D_J$ (eq. 25), and the mechanical energy of the companion's
jets relative to the envelope binding energy $D_{\rm acc}$ (eq. 26).
}
\end{figure}
with the same scaling as used in equations
(22), (25), and (26), respectively.
I note the following:
(1) The three functions depend in the same way on
$M_2/M_{\rm env}$.
(2) If $D_J \gtrsim 1-3$ and the giant's radius does
not expand to the initial orbital separation, the system will not
enter a CE phase; the exact value depends on how much angular
momentum is removed by the wind.
Hence a CE with giant at late stages requires
$M_2 \lesssim M_{\rm env}$.
(3) The deposition of angular momentum by itself does not
remove the envelope. But rotating giants may be more efficient in
utilizing the luminosity to expel the envelope, e.g.,
by forming more dust.
(4) If $D_J$ is larger, the spiraling-in takes more time,
because there is a need to remove angular momentum from the system.
This increases $\tau_d$, hence $D_{\rm acc}$ becomes larger.
(5) In most cases the orbital energy deposited into the envelope
becomes significant only when the companion is deep in the envelope.
At very small orbital separation the companion may loss mass to
the giant's core (e.g., Ivanova, Podsiadlowski \& Spruit 2002),
releasing more gravitational energy.
(6) At early stages, deposition of angular momentum, which may help
forming dust, and the mechanical energy released by the accreting
companion plays a larger role than the orbital energy is
in expelling the envelope (in addition to the `regular' RGB and
AGB wind).
This may end with large post-CE orbital separation
(Soker \& Harpaz 2003). 

% ===================================================
\subsection{Accumulated Effects} 
% ===================================================

A more appropriate indicator for the significant of the different
processes is their accumulated effect.
I repeat here the treatment of the previous subsection,
but with the total deposited energy and angular momentum.
The total energy deposited by the companion as it
spirals-in from initial orbital separation $a_0$ to $a$ is 
\begin{equation}
\Delta E_{\rm orb} = \frac {G M_1 M_2} {2 a} \left(1-\frac{a}{a_0}\right) .
\end{equation}
Again, a fraction $1-\alpha_e$ of this energy be radiated away.
The total relative energy deposition as the companion spirals-in,
the energy factor, is defined as
\begin{equation}
A_E (M_2,a) \equiv \frac {\alpha_e \Delta E_{\rm orb}}{\Delta E_{\rm bind}}
= \frac {\alpha_e}{2 B_{\rm env}} \frac {M_2}{M_{\rm env}}
\frac {R_\ast}{a} \left(1-\frac{a}{a_0}\right).
\end{equation}
The orbital angular momentum deposited to the envelope as the
orbit shrinks is given, for $M_1 \gg M_2$, by
\begin{equation}
\Delta  J_o
=  ( G M_1 a_0 )^{1/2} M_2
\left[1- \left( \frac{a}{a_0}\right)^{1/2} \right]. 
\end{equation}
I assume that angular momentum deposition starts
with tidal interaction, when $a_0 \sim 4 R_\ast$, and
use this value for $a_0$. 
The total angular momentum depostion factor is defined by 
\begin{equation}
A_J \equiv \frac {\Delta J_o}{J_{\rm env} ({\rm max})}
= \frac {1}{k_e} \frac{M_2}{M_{\rm env}}
\left[1- \left( \frac{a}{a_0}\right)^{1/2} \right] .
\end{equation}
The ratio of the angular momentum factor to the energy factor is
\begin{equation}
\frac {A_J}{A_E} = 100
\left( \frac {k_e}{0.2} \right)^{-1}
\left( \frac {\alpha_e}{0.5} \right)^{-1}
\left( \frac {B_{\rm env}}{5} \right)
\frac{a}{R_\ast}
\left[1 + \left( \frac{a}{a_0}\right)^{1/2} \right]^{-1} .
\end{equation}

For the parameters used to scale the last equation, it turns out
that energy deposition dominates over angular momentum
deposition only when $a \lesssim 0.01 R_\ast$.
For a giant of $R_\ast \sim 1 \AU$, this occurs when $a \sim 2 R_\odot$.
By then many companions will go through a RLOF process.
My conclusion is that for the mass loss process, in most cases
it is angular momentum deposition which causes large effects.
This is true mainly in giants which have high mass loss rate,
such that the rotating envelope will facilitate much higher mass
loss rate, e.g., by enhancing dust formation.

% ===================================================
\section{SUMMARY} 
% ===================================================

The main goal of the present paper is to point to the
caution one must take in using the $\alpha_{\rm CE}$
parameter when studying CE evolution.
Namely, the orbital energy deposited to the giant's envelope
is not always the main effect leading, directly or indirectly, to
the removal of the envelope.
For that I considered here the deposition of energy from
the accreting companion and the deposition of
orbital angular momentum to the giant's envelope. 
The main results can be summarized as follows.
\begin{enumerate}
\item When inside the envelope of a giant,
    a main sequence companion is unlikely to blow jets, or a 
    collimated fast wind (CFW, i.e., less collimated jets),
    or it will marginally do so only when in the outer
    parts of the envelope.
\item A WD companion is more likely to blow jets or a CFW
\item These jets, even if exist, whether from a WD or a MS
    companion, are not likely to exit the envelope at a high
    speed during the CE phase. Hence, they are not likely
    to play a major role in shaping the circumbinary matter.
    Jets might be blown by the companion before entering the
    CE (Soker \& Rappaport 2000), or one or two of the stars after the
    CE ends (Soker \& Livio 1994).
    This explains the observations that PNs with
    binary nuclei are not bipolar PNs, i.e., have no lobes,
    beside NGC 2346, with the longest known orbital period.
    I do expect that some binary progenitors of bipolar PNs entered 
    the CE phase at late stages, and that now the orbital separation
    is $\sim 0.1-1 \AU$. These systems are hard to detect 
    (Bond 2000). 
    To obtain a quantitative result, the CE population synthesis calculations 
    of Yungelson, Tutukov, \& Livio (1993) should be repeated but with 
    enhanced mass loss rate from rotating AGB stars included. 
\item The CFW or jets, if exist, may inflate a 
    bubble (with a complicated structure because of the orbital
    motion), hence playing a significant role in expelling
    the outer layers of the envelope when the companion
    is still orbiting in the outer envelope region.
\item In many cases the effects due to angular momentum deposition into 
    the envelope seem more influential in removing the envelope than
    orbital energy deposition, assuming that fast rotating 
    envelopes have high mass loss rates.
    This is true for stellar as well as substellar companions.
    The energy source is the giant luminosity due
    to nuclear energy production in the core.
    The Eddington luminosity of an accreting stellar companion is of 
    the order of the giant's luminosity, and can farther increase the 
    mass loss rate (Iben \& Livio 1993; Armitage \& Livio 2000).
\item My results here iterate earlier claims (Soker 2002; 
    Soker \& Harapz 2003) that a high degree of cautious should 
    be taken when applying the $\alpha_{\rm CE}$ parameter for 
    the removal of CEs.
    For example, the conclusions of some papers that another 
    energy source, e.g., ionization energy of the envelope, 
    is required to remove the envelope (see criticism in 
    Soker \& Harpaz 2003) are questionable.
\end{enumerate}
                                                                 
\acknowledgements
I thank Mario Livio for very helpful and detailed comments
at the beginning of this project. 
This research was partially supported by the
Israel Science Foundation.

\end{document}